\documentstyle[prb,aps]{revtex}

\begin{document}
\date{\today}
\draft
\title{Resonant Tunneling through Linear Arrays of Quantum Dots}
\author{M. R. Wegewijs and Yu. V. Nazarov}
\address{Faculty of Applied Sciences and \\
Delft Institute of Microelectronics and Submicrontechnology (DIMES), \\
Delft University of Technology, 2628 CJ Delft, The Netherlands}
\maketitle

\begin{abstract}
We theoretically investigate resonant tunneling through a linear array of
quantum dots with subsequent tunnel coupling. We consider two limiting
cases: (i) strong Coulomb blockade, where only one extra electron can be
present in the array (ii) limit of almost non-interacting electrons. We
develop a density matrix description that incorporates the coupling of the
dots to reservoirs. We analyze in detail the dependence of the stationary
current on the electron energies, tunnel matrix elements and rates, and on
the number of dots. We describe interaction and localization effects on the
resonant current. We analyze the applicability of the approximation of
independent conduction channels. We find that this approximation is not
valid when at least one of the tunnel rates to the leads is comparable to
the energy splitting of the states in the array. In this case the
interference of conduction processes through different channels suppresses
the current.
\end{abstract}

\section{Introduction}

In recent years arrays of quantum dots have received an increasing amount of
interest. With the progress of fabrication techniques quantum dot arrays are
coming within the reach of experimental investigation \cite{bib:Dots}. If
the electron levels in the individual dots are aligned we encounter here a
situation of resonant tunneling. In this regime, the transport in the array
becomes sensitive to precise matching of the electron levels in the dots
that can be controlled by external gates. This opens up new possibilities to
control the transport and perform sensitive measurements even in the
simplest case of two dots \cite{bib:Exp}.

Resonant tunneling in arrays of quantum dots and layered semiconductor
heterostructures exhibit some similarities. The latter situation has been
intensively studied in the context of possible Bloch oscillations \cite
{bib:Hetero}. However, Coulomb blockade dominates the properties of the
arrays of quantum dots so that electron-electron interaction cannot be
neglected as in the case of layered heterostructures \cite
{bib:KorotkovAverinLikharev}, \cite{bib:FrishmanGurvitz}. A way to
circumvent this difficulty is to perform an exact diagonalization of
electron states in the array\ of coupled dots. Then one considers
independent tunneling transitions between the resulting many-electron states 
\cite{bib:ChenKlimeckDatta}. This we call the independent channel
approximation. This is approximate because it disregards the simultaneous
tunneling an electron through multiple conduction channels. Another approach
is to restrict the basis to the resonant states of the uncoupled dots. Then
the tunneling between the dots and the reservoirs is incorporated into a
modified Liouville equation for the density matrix in this basis. For two
quantum dots this has been done in\cite{bib:Nazarov} and here we extend this
approach to the case of an array of an arbitrary number of dots.

In this paper we concentrate on an array of quantum dots where dots are
connected in series and a tunnel coupling exists only between neighboring
dots. This is the most interesting case because there is a unique path for
the current and changes in any dot strongly affect the transport through the
whole array. The first and last dot of the array are connected to leads. We
assume that the voltage bias is sufficiently high so that the energy change
during the tunneling of an electron between a reservoir and the array is
much larger than the energy uncertainty due to this tunneling. We also
assume that the resonant electronic energies in the array lie well between
the Fermi levels of the leads. This enables us to use the density matrix
approach. We consider two limiting cases of the electron-electron
interaction within the array. In the first case we assume that the long
range Coulomb repulsion between electrons in different dots of the array is
so strong so that only one or no extra electrons are present in the array
(Coulomb blockade). This is to be contrasted with the case of ``free''
electrons. As we explain below (Section \ref{sec:Arrays}) we do not
disregard interactions completely in the latter case but rather account only
for strong repulsion within each dot.

Using the density matrix approach in the basis of localized states we have
obtained analytical results for the stationary current. Our results hold for
arbitrary values of the parameters (within the applicability of our model)
characterizing the array like dot energies and tunnel couplings: no
assumption about homogeneity of the array has been made. This may facilitate
the comparison with experiments and the design of resonant tunneling
devices. We report the effects of localization and Coulomb repulsion on the
resonant current when the energy level of the first and the last dot are
independently varied. We have also considered another picture of the
transport using the approximation of independent conduction channels in the
array of dots. Using the density matrix approach in the basis of delocalized
states we have calculated the occupations of the channels and their
contributions to the current. We discuss in detail the range of validity of
this approximation. In the limit of both weak {\em and} strong coupling to
one or both of the leads we obtain results in agreement with the former more
general calculations. However, there can be substantial deviations from the
predictions of this model when the tunnel rates and the coherent interdot
couplings are comparable. To illustrate this we study the dependence of the
current on the transparencies of the tunnel barriers and find unusual
features due to the interference of electrons during tunneling.

The outline of the paper is as follows: in Section \ref{sec:Arrays} we
introduce the density matrix description of a multidot system coupled to
leads which we apply to the Coulomb blockade case in Section \ref{sec:CB}
and to the ``free'' electron case\ in Section \ref{sec:F}. In Section \ref
{sec:Appr} we compare the results with those obtained from the independent
channel approximation and we discuss the deviations.\ We formulate our
conclusions in Section \ref{sec:Concl}.

\section{Array of quantum dots coupled to reservoirs}

\label{sec:Arrays}Let's first consider an array of $N$ quantum dots (the
``device'') without any contacts (Fig. \ref{fig:Ndots}).We consider a
quantum dot as some complex system with discrete many-electron states of
which only two ground states participate in resonant transport. The one
state is related to the other by the addition of an {\em extra} electron
costing an addition energy which includes intradot charging i.e. the Coulomb
interactions between electrons in the dot. (For simplicity we disregard the
electron's spin degrees of freedom\cite{bib:SpinDegeneracy}). We label the
resonant states of dot $i=1,\ldots ,N$ by the number of extra electrons $%
n_{i}=0,1$ and introduce fermionic operators $\hat{a}_{i}^{\dagger },\hat{a}%
_{i}$ which create resp. annihilate an extra electron in dot $i$. The
many-electron eigenstates of the array of {\em uncoupled} dots will be
denoted by $|n_{1}\ldots n_{N}\rangle =|\{n_{k}\}\rangle =\prod_{k=1}^{N}%
\hat{a}_{k}^{\dagger n_{k}}|\left\{ 0\right\} \rangle $. Let $\varepsilon
_{i}$ denote the energy for adding an electron to dot $i$ and let $u_{\left|
i-j\right| }\geq 0$ be the interdot charging energy due to the repulsive
Coulomb interaction between the pair of extra electrons in dots $i\neq j$. A
coherent interdot coupling with matrix element $t_{i}=t_{i}^{\ast }$
accounts for the tunneling of electrons between dots $i$ and $i+1$. We
obtain a Hubbard-type Hamiltonian $\hat{H}=\hat{H}_{0}+\hat{H}_{u}$ for the
array of {\em coupled} dots:%
\begin{mathletters}%
%
\begin{eqnarray}
\hat{H}_{0} &=&\sum\limits_{i=1}^{N}\varepsilon _{i}\hat{a}_{i}^{\dagger }%
\hat{a}_{i}+\sum\limits_{i=1}^{N-1}t_{i}\left( \hat{a}_{i+1}^{\dagger }\hat{a%
}_{i}+\hat{a}_{i}^{\dagger }\hat{a}_{i+1}\right)  \label{eq:H0} \\
\hat{H}_{u} &=&\sum\limits_{i<j=2}^{N}u_{\left| i-j\right| }\hat{a}%
_{i}^{\dagger }\hat{a}_{i}\hat{a}_{j}^{\dagger }\hat{a}_{j}  \label{eq:HU}
\end{eqnarray}
\label{eq:H}%
\end{mathletters}%
%
The density operator of the device $\hat{\sigma}$ evolves according to the
Liouville equation $\partial _{t}\hat{\sigma}=-$i$[\hat{H},\hat{\sigma}]$ ($%
\hbar \equiv 1$) which conserves probability i.e. $%
\mathop{\rm Tr}%
\hat{\sigma}\left( t\right) =1$. By expanding $\hat{\sigma}$ in the
many-electron eigenstates of the uncoupled dots we obtain a $2^{N}\times
2^{N}$ density-matrix $\mbox{\boldmath $ \sigma $}=(\sigma
_{\{n_{k}\},\{n_{k}^{\prime }\}})$: 
\[
\hat{\sigma}=\sum_{\{n_{k}\},\{n_{k}^{\prime }\}}\sigma
_{\{n_{k}\},\{n_{k}^{\prime }\}}|\{n_{k}\}\rangle \langle \{n_{k}^{\prime
}\}| 
\]
We additionally introduce an $N\times N$ Hermitian matrix $%
\mbox{\boldmath $
\rho $}$ with expectation values of single-electron operators\ ($%
i,j=1,\ldots ,N$): 
\[
\rho _{ij}=\langle \hat{a}_{j}^{\dagger }\hat{a}_{i}\rangle =%
\mathop{\rm Tr}%
\hat{a}_{j}^{\dagger }\hat{a}_{i}\hat{\sigma} 
\]
Using the fermionic commutation relations we find%
\begin{mathletters}%
%
\begin{eqnarray}
\rho _{ii} &=&\sum\limits_{\{n_{k\neq i}\}}\sigma _{n_{1}\ldots 
\mathrel{\mathop{1}\limits_{i}}%
\ldots n_{N},n_{1}\ldots 
\mathrel{\mathop{1}\limits_{i}}%
\ldots n_{N}}  \label{eq:RhoDefii} \\
\rho _{ij} &=&\sum\limits_{\{n_{k\neq i,j}\}}\left( -1\right)
^{n_{i+1}+\ldots +n_{j-1}}\sigma _{n_{1}\ldots 
\mathrel{\mathop{1}\limits_{i}}%
\ldots 
\mathrel{\mathop{0}\limits_{j}}%
\ldots n_{N},n_{1}\ldots 
\mathrel{\mathop{0}\limits_{i}}%
\ldots 
\mathrel{\mathop{1}\limits_{j}}%
\ldots n_{N}}  \label{eq:RhoDefij}
\end{eqnarray}
\label{eq:RhoDef}%
\end{mathletters}%
%
where $\{n_{k\neq i,j}\}$ indicates that we sum over $n_{k}=0,1$ for all
dots $k=1,\ldots ,N$ except $i,j$. We will refer to $\mbox{\boldmath $ \rho$}
$ as the average occupation matrix w.r.t. electrons in individual dots
whereas $\mbox{\boldmath $ \sigma $}$ is the probability-density matrix
w.r.t. many-electron states of the array.

Now we include leads $L$ and $R$ connected to resp. the first and last dot
of the device by a tunnel barrier (Fig. \ref{fig:Ndots}). The leads are
considered here as electron reservoirs at zero temperature with a continuum
of states filled up to their resp. Fermi levels $\mu _{L}$ and $\mu _{R}$.
We assume that the energy levels of the device are located well between the
chemical potentials of the leads i.e. $\mu _{L}\gg \varepsilon _{i},t_{i}\gg
\mu _{R}$ (large bias) and that the level widths are much smaller than the
bias i.e. $\mu _{L}-\mu _{R}\gg \Gamma _{L,R}$ (discrete states).{\bf \ }%
Under these assumptions an evolution equation for the density-matrix $%
\mbox{\boldmath $ \sigma $}$ of the device can be derived \cite
{bib:GurvitzPrager},\cite{bib:GurvitzPRB57} by incorporating the details of
the lead states into tunnel rates. Due to the high bias electrons only
tunnel through the barrier from lead $L$ to dot $1$ with a rate $\Gamma
_{L}=2\pi D_{L}\left| t_{L}\right| ^{2}$and from dot $N$ to lead $R$ with
rate $\Gamma _{R}=2\pi D_{R}\left| t_{R}\right| ^{2}$. We assume that the
density of states $D_{L,R}$ in leads is constant and that the tunneling
matrix elements $t_{L,R}$ between lead and dot states depend only weakly on
the energy. Due to the destructive interference of an electron tunneling
between a discrete state in the device and the continuum of states in a
reservoir the rate for tunneling in one direction ($L\rightarrow
1,N\rightarrow R$) is constant in time whereas the rate of the reversed
process ($L\leftarrow 1,N\leftarrow R$) is zero\cite{bib:Merzbacher}. In
general transitions between discrete many-electron states $a,b$ of a device
with Hamiltonian $\hat{H}$ induced by tunneling to and from reservoirs can
be included into a modification of the Liouville equation\cite
{bib:GurvitzPRB57}: 
\begin{eqnarray}
\partial _{t}\sigma _{ab} &=&-\text{i}[\hat{H},\hat{\sigma}]_{ab}  \nonumber
\\
&&-\frac{1}{2}(\sum_{a^{\prime }}\Gamma _{a\rightarrow a^{\prime
}}+\sum_{b^{\prime }}\Gamma _{b\rightarrow b^{\prime }})\sigma
_{ab}+\sum_{a^{\prime }b^{\prime }}\Gamma _{ab\leftarrow a^{\prime
}b^{\prime }}\sigma _{a^{\prime }b^{\prime }}  \label{eq:Sigma}
\end{eqnarray}
The first term of the modification describes the {\em separate} decay of
states $a\rightarrow a^{\prime }$ and $b\rightarrow b^{\prime }$ due to (in
general) {\em different} tunneling events between the device and the
reservoir with resp. rates $\Gamma _{a\rightarrow a^{\prime }}$ and $\Gamma
_{b\rightarrow b^{\prime }}$. The second term describes the {\em joint}
generation of states $a$ $\leftarrow a^{\prime }$ and $b\leftarrow b^{\prime
}$ due to a {\em single} tunneling event occurring with rate $\Gamma
_{ab\leftarrow a^{\prime }b^{\prime }}$: the coherence lost by the
simultaneous decay of states $a^{\prime },b^{\prime }$ is transferred to
states $a,b$. When we assume that there is a unique path for the current,
then each state $a$ is generated from a unique state $a^{\prime }$ by the
tunneling of an electron to or from a reservoir: $\Gamma _{aa\leftarrow
a^{\prime }b^{\prime }}=\Gamma _{a\leftarrow a^{\prime }}\delta _{a^{\prime
}b^{\prime }}$. The modified Liouville equation conserves probability:
summing the equations for the diagonal elements gives $\partial _{t}%
\mathop{\rm Tr}%
\hat{\sigma}\left( t\right) =0$ so 
\begin{equation}
\sum_{\{n_{k}\}}\sigma _{\{n_{k}\},\{n_{k}\}}=1  \label{eq:SigmaCons}
\end{equation}

We consider the following two cases for resonant transport through the
array: (i) ``Free'' electron (F) case where interdot Coulomb repulsion is
negligible i.e. $u_{\left| i-j\right| }\equiv 0$; up to $N$\ electrons can
populate the array and all $2^{N}$\ many-electron states lie between the
chemical potentials $\mu _{L}\gg \mu _{R}$ and participate in resonant
transport. (ii) Coulomb blockade (CB) case where the interdot Coulomb
repulsion is so strong that the smallest charging energy is large relative
to the bias i.e. $u_{N-1}\gg \mu _{L}-\mu _{R}$; many-electron states with
more than one electron are highly improbable and can be neglected for
resonant transport. We point out that in general the presence of electrons
in the array modifies the rates for tunneling to or from the leads by
Coulomb repulsion\cite{bib:GurvitzPrager}. However, in the limiting cases
considered here there is either no repulsion (F) or no other electron
present in the array (CB) so two parameters $\Gamma _{L}$ and $\Gamma _{R}$
suffice to incorporate the details of the electronic states in the resp.
leads.{\em \ }The current flowing from dot $N$ to reservoir $R$ is
determined by the average occupation and the tunnel rate: 
\begin{equation}
\frac{I_{N}\left( t\right) }{e}=\Gamma _{R}\rho _{NN}\left( t\right)
\label{eq:I}
\end{equation}
With the leads included the dynamics of the average occupation matrix $%
\mbox{\boldmath $ \rho$}$ should be calculated from the full density matrix $%
\mbox{\boldmath $ \sigma $}$\ which evolves according to an equation of the
type (\ref{eq:Sigma}). However, in both cases considered here we can derive
a dynamical equation for the matrix $\mbox{\boldmath $ \rho$}$\ which is
solved more easily.

\section{Current in Coulomb blockade case}

\label{sec:CB}When the long range Coulomb repulsion is so strong that at
most one electron can be present in the array of dots we can restrict the
set of many-electron basis states to $|0\ldots 0\rangle ,|0\ldots 
\mathrel{\mathop{1}\limits_{i}}%
\ldots 0\rangle ,i=1,\ldots ,N$. The average occupations of individual dots
are simply equal to the non-zero probability densities of these states and 
\[
\rho _{ij}=\sigma _{0\ldots 
\mathrel{\mathop{1}\limits_{i}}%
\ldots 0,0\ldots 
\mathrel{\mathop{1}\limits_{j}}%
\ldots 0} 
\]
Conservation of probability (\ref{eq:SigmaCons}) suggests that we
additionally define an average occupation of the many-electron vacuum state: 
\[
\rho _{00}=\sigma _{0\ldots 0,0\ldots 0} 
\]
This quantity is positive $0\leq \rho _{00}\leq 1$ and satisfies a
conservation law: 
\begin{equation}
\rho _{00}+\sum_{k=1}^{N}\rho _{kk}=1  \label{eq:RhoCBCons}
\end{equation}
In the restricted basis the matrix elements of the Hamiltonian $\hat{H}$
which describe the coherent part of the evolution of the state only involve $%
\hat{H}_{0}$ (eq. (\ref{eq:H0})). Modifying the Liouville equation according
to (\ref{eq:Sigma}) we obtain a dynamical equation for the average
occupation matrix%
\begin{mathletters}%
%
\begin{eqnarray}
\text{$\partial _{t}\rho $}_{00} &=&-\Gamma _{L}\rho _{00}+\text{$\Gamma
_{R}\rho _{NN}$}  \label{eq:RhoCB00} \\
\text{$\partial _{t}\rho $}_{ii} &=&\text{i}\left( t_{i-1}\rho
_{ii-1}+t_{i}\rho _{ii+1}-t_{i-1}\rho _{i-1i}-t_{i}\rho _{i+1i}\right) 
\nonumber \\
&&+\text{$\Gamma _{L}\rho _{00}\delta _{i1}$}-\text{$\Gamma _{R}\rho
_{NN}\delta _{iN}$}  \label{eq:RhoCBii} \\
\partial _{t}\rho _{ij} &=&\text{i}\left( \varepsilon _{j}-\varepsilon
_{i}\right) \rho _{ij}  \nonumber \\
&&+\text{i}\left( t_{j-1}\rho _{ij-1}+t_{j}\rho _{ij+1}-t_{i-1}\rho
_{i-1j}-t_{i}\rho _{i+1j}\right)  \nonumber \\
&&-\frac{1}{2}\Gamma _{R}\rho _{iN}\delta _{jN}  \label{eq:RhoCBij}
\end{eqnarray}
\label{eq:RhoCB}%
\end{mathletters}%
%
where $j>i=1,\ldots ,N$ and $t_{0}=t_{N}\equiv 0$. In the rhs of (\ref
{eq:RhoCB00}) the negative contribution describes the decay of the vacuum
state due to the tunneling of an electron from lead $L$ to dot $1$ with rate 
$\Gamma _{L}$ whereas the positive contribution describes the generation of
this state due to the tunneling of the (only) electron in the device from
dot $N$ to lead $R$. In (\ref{eq:RhoCBij}) there is only a negative
contribution due to the tunneling of an electron {\em out} of dot $N$ to
reservoir $R$ with rate $\Gamma _{R}.$ There is no negative contribution
with $\Gamma _{L}$.because we have incorporated Coulomb blockade: we
disregard the decay of many-electron states with $1$ electron to a state
with $2$ electrons which occurs when an electron tunnels from lead $L$ {\em %
into} dot $1$. We can eliminate $\rho _{00}$ from eqns. (\ref{eq:RhoCB})
using eq. (\ref{eq:RhoCBCons}) and obtain $N^{2}$ equations for the average
occupations.

Eqns. (\ref{eq:RhoCB}) can be used to describe non-stationary transport with
a typical relaxation time scale $\Gamma _{L,R}^{-1}$. Here we are interested
in the stationary limit $\partial _{t}\mbox{\boldmath $
\rho$}=0$ only. In general the solution of eq. (\ref{eq:RhoCB}) can be
obtained by inverting a matrix of dimension $N^{2}$. However, since the
system under consideration is an array with subsequent tunnel coupling most
of the equations only relate matrix elements of $\mbox{\boldmath $
\rho$}$ on neighboring dots. One can obtain the solution by iteratively
expressing all matrix elements in terms of $\rho _{00}$ and finally making
use of the normalization constraint (\ref{eq:RhoCBCons}). The resulting
stationary current in general reads as 
\begin{equation}
\frac{I_{N}^{\text{CB}}}{e}=\frac{1}{%
{\displaystyle{1 \over \Gamma _{L}}}%
+%
{\displaystyle{1 \over \Gamma _{R}}}%
F_{N}+%
{\displaystyle{\Gamma _{R} \over 4t_{N-1}^{2}}}%
F_{N-1}}  \label{eq:ICB}
\end{equation}
Here dimensionless expression $F_{N}$\ depends only on $\varepsilon
_{1},\ldots \varepsilon _{N}$ through the differences $\varepsilon
_{ij}\equiv \varepsilon _{i}-\varepsilon _{j},i<j$ and on $t_{1}\ldots
t_{N-1}$. Note the curious property of eq. (\ref{eq:ICB}): $F_{N}$ enters
the expressions for both $I_{N}$ and $I_{N+1}$. This helps when deriving
expressions for the current through an array with an increasing number of
dots. For $N=2$ we reproduce the result of Stoof and Nazarov\cite
{bib:StoofNazarov}: 
\begin{equation}
\frac{I_{2}^{\text{CB}}}{e}=\frac{1}{%
{\displaystyle{1 \over \Gamma _{L}}}%
+%
{\displaystyle{1 \over \Gamma _{R}}}%
\left( 2+\left( 
{\displaystyle{\varepsilon _{12} \over t_{1}}}%
\right) ^{2}\right) +%
{\displaystyle{\Gamma _{R} \over 4t_{1}^{2}}}%
}  \label{eq:ICB2}
\end{equation}
For $N=3$ we obtain for the current through a triple dot system 
\begin{equation}
\frac{I_{3}^{\text{CB}}}{e}=\frac{1}{%
{\displaystyle{1 \over \Gamma _{L}}}%
+%
{\displaystyle{1 \over \Gamma _{R}}}%
\left( 3+%
{\displaystyle{\varepsilon _{12}\varepsilon _{13}+2t_{1}^{2}-t_{2}^{2} \over t_{1}t_{2}}}%
{\displaystyle{\varepsilon _{12}\varepsilon _{13}+t_{1}^{2}-t_{2}^{2} \over t_{1}t_{2}}}%
+%
{\displaystyle{\varepsilon _{12}+2\varepsilon _{23} \over t_{2}}}%
{\displaystyle{\varepsilon _{12}+\varepsilon _{23} \over t_{2}}}%
\right) +%
{\displaystyle{\Gamma _{R} \over 4t_{2}^{2}}}%
\left( 2+%
{\displaystyle{\varepsilon _{12} \over t_{1}}}%
\right) }  \label{eq:ICB3}
\end{equation}
One can show $F_{N}\left( \varepsilon _{1},\ldots \varepsilon
_{N},t_{1}\ldots t_{N-1}\right) =N$ for $\varepsilon _{ij}\ll t_{i}=t$. This
property will be used later on and it is derived in an other way in Section 
\ref{sec:Weak}.

We consider eq. (\ref{eq:ICB}) in more detail for equal interdot couplings $%
t_{i}=t$\ and several different configurations of the levels $\varepsilon
_{i}$. At resonance $\varepsilon _{ij}=0$ we find 
\[
\rho _{ii}=\frac{%
{\displaystyle{1 \over \Gamma _{R}}}%
+%
{\displaystyle{\Gamma _{R} \over 4t^{2}}}%
\left( 1-\delta _{iN}\right) }{%
{\displaystyle{1 \over \Gamma _{L}}}%
+%
{\displaystyle{1 \over \Gamma _{R}}}%
N+%
{\displaystyle{\Gamma _{R} \over 4t^{2}}}%
\left( N-1\right) }
\]
and the current is maximal: 
\begin{equation}
{\displaystyle{I_{N,\text{max}}^{\text{CB}} \over e}}%
=\frac{1}{%
{\displaystyle{1 \over \Gamma _{L}}}%
+%
{\displaystyle{1 \over \Gamma _{R}}}%
N+%
{\displaystyle{\Gamma _{R} \over 4t^{2}}}%
\left( N-1\right) }  \label{eq:ImaxCB}
\end{equation}
Clearly the current is reduced when the array increases in size: the number
of states participating in transport relative to the number of states of the
array decreases $\propto 1/N$ due to the Coulomb blockade.{\bf \ }In an
actual system we expect a positive{\bf \ }deviation from this decrease to
occur when the spatial size of the array exceeds some range over which the
Coulomb repulsion cannot exclude the occupation of a second dot in the
array. Away from resonance i.e. $\varepsilon _{ij}=O\left( \varepsilon
\right) \gg t$ the conduction of the device decays exponentially with the
size of the array $N$ due to the localization of {\em the }electron in one
of the individual dots: $I_{N}^{\text{CB}}/e\propto \Gamma _{R}(\varepsilon
/t)^{-2(N-1)}$. To illustrate this we vary only the last level (Fig. \ref
{fig:ICB}a) i.e. we consider the solution of (\ref{eq:RhoCB}) for $%
\varepsilon _{i}=\varepsilon _{N}\delta _{iN}$: 
\[
\frac{I_{N}^{\text{CB}}}{e}=\frac{1}{%
{\displaystyle{1 \over \Gamma _{L}}}%
+%
{\displaystyle{1 \over \Gamma _{R}}}%
\left( N+\left( N-1\right) \left( 
{\displaystyle{\varepsilon _{N} \over t}}%
\right) ^{2}\right) +%
{\displaystyle{\Gamma _{R} \over 4t^{2}}}%
\left( N-1\right) }
\]
A{\bf s} we increase the number of dots the curve keeps its Lorentzian shape
w.r.t. $\varepsilon _{N}$ and always depends on both $t$\ and $\Gamma _{R}$
since the electron is localized only just before tunneling out of the array: 
\[
{\displaystyle{I_{N}^{\text{CB}} \over I_{N,\text{max}}^{\text{CB}}}}%
\stackrel{N\rightarrow \infty }{=}%
{\displaystyle{1+\left( %
{\displaystyle{\Gamma _{R} \over 2t}}\right) ^{2} \over 1+\left( %
{\displaystyle{\Gamma _{R} \over 2t}}\right) ^{2}+\left( %
{\displaystyle{\varepsilon _{N} \over t}}\right) ^{2}}}%
\]
Now we vary only the first level (Fig. \ref{fig:ICB}b) i.e. we consider the
solution of (\ref{eq:RhoCB}) for $\varepsilon _{i}=\varepsilon _{1}\delta
_{i1}$: 
\[
\frac{I_{N}^{\text{CB}}}{e}=\frac{1}{%
{\displaystyle{1 \over \Gamma _{L}}}%
+%
{\displaystyle{1 \over \Gamma _{R}}}%
\left( N+\left( \left( 
{\displaystyle{\varepsilon _{1} \over t}}%
\right) ^{2}+\left( 
{\displaystyle{\Gamma _{R} \over 2t}}%
\right) ^{2}\right) \sum\limits_{k=1}^{N-1}k\left( 
{\displaystyle{\varepsilon _{1} \over t}}%
\right) ^{2\left( N-1-k\right) }\right) }
\]
For large $N$ the normalized current vanishes when the detuning exceeds the
tunnel coupling (since an electron is localized in the first dot) whereas
near resonance the peak takes on a parabolic shape which is independent of
the tunnel rate $\Gamma _{R}$: 
\[
{\displaystyle{I_{N}^{\text{CB}} \over I_{N,\text{max}}^{\text{CB}}}}%
\stackrel{N\rightarrow \infty }{=}\left\{ 
\begin{array}{cc}
1-\left( 
{\displaystyle{\varepsilon _{1} \over t}}%
\right) ^{2} & \varepsilon _{1}/t<1 \\ 
0 & \varepsilon _{1}/t>1
\end{array}
\right. 
\]
For the case where the energies are configured as a ``Stark ladder'' of
total width $\varepsilon $ we have plotted\ the current in Fig. \ref
{fig:IStark}b. The localization of electrons clearly dominates the current
since the tails of the current peak decrease rapidly with increasing $N$\ as
in Fig. \ref{fig:ICB}a.

\section{Current in ``free'' electron case}

\label{sec:F}When interdot Coulomb repulsion is altogether disregarded all $%
2^{N}$ many-electron states $|\{n_{k}\}\rangle $ of the array of dots must
be taken into account. Modifying the Liouville equation with $\hat{H}=\hat{H}%
_{0}$ according to the general prescription (\ref{eq:Sigma}) we obtain the
following set of $2^{2N}$ equations for the density matrix:%
\begin{mathletters}%
%
\begin{eqnarray}
\partial _{t}\sigma _{n_{1}\ldots n_{N},n_{1}^{\prime }\ldots n_{N}^{\prime
}} &=&-\text{i}\left[ \hat{H}_{0},\sigma \right] _{n_{1}\ldots
n_{N},n_{1}^{\prime }\ldots n_{N}^{\prime }}  \label{eq:SigmaFt} \\
&&-%
{\displaystyle{1 \over 2}}%
\Gamma _{L}(\left( 1-n_{1}\right) +\left( 1-n_{1}^{\prime }\right) )\sigma
_{n_{1}\ldots n_{N},n_{1}^{\prime }\ldots n_{N}^{\prime }}-%
{\displaystyle{1 \over 2}}%
\Gamma _{R}(n_{N}+n_{N}^{\prime })\sigma _{n_{1}\ldots n_{N},n_{1}^{\prime
}\ldots n_{N}^{\prime }}  \label{eq:SigmaF-} \\
&&+\Gamma _{L}n_{1}n_{1}^{\prime }\sigma _{0n_{2}\ldots n_{N},0n_{2}^{\prime
}\ldots n_{N}^{\prime }}+\Gamma _{R}(1-n_{N})(1-n_{N}^{\prime })\sigma
_{n_{1}\ldots n_{N-1}1,n_{1}^{\prime }\ldots n_{N-1}^{\prime }1}
\label{eq:SigmaF+}
\end{eqnarray}
\label{eq:SigmaF}%
\end{mathletters}%
%
Here (\ref{eq:SigmaFt}) describes transitions between the non-orthogonal
states of the device with a fixed number of electrons due to the tunneling
between neighboring dots. In contrast to the Coulomb blockade case both
tunneling into ($\Gamma _{L}$) {\em and} out of the array ($\Gamma _{R}$)
give negative contributions (\ref{eq:SigmaF-}). Furthermore, there are
tunnel processes which induce a transition between 2 pairs of many-electron
states and give a positive contribution (\ref{eq:SigmaF+}) to the
coherences. Only a subset of (\ref{eq:SigmaF}) forms a closed system of
equations for the diagonal and some non-diagonal elements of $%
\mbox{\boldmath $
\sigma$}$ (The remaining equations only couple a closed set of non-diagonal
elements which are irrelevant). From this subset we can derive an even
simpler dynamical equation for the average occupation matrix $%
\mbox{\boldmath $
\rho$}$. Let's first derive the coherent part of this equation by
considering ``free'' electrons in the array of dots without the leads.
Because the Hamiltonian consist of only one-electron operators ($%
H_{0}=\sum_{kl=1}^{N}H_{0kl}a_{k}^{\dagger }a_{l}$), the commutator in the
Heisenberg eq. of motion for $\rho _{ij}=\langle a_{j}^{\dagger
}a_{i}\rangle $ is readily expressed in other one-electron operators and the
matrix elements of $\hat{H}_{0}$: 
\begin{eqnarray}
\text{i}\partial _{t}\rho _{ij} &=&\langle \left[ a_{j}^{\dagger }a_{i},\hat{%
H}_{0}\right] \rangle  \nonumber \\
&=&\sum_{l=1}^{N}H_{0il}\langle a_{j}^{\dagger }a_{l}\rangle
-\sum_{k=1}^{N}\langle a_{k}^{\dagger }a_{i}\rangle H_{0kj}=\left[ {\bf H}%
_{0},\mbox{\boldmath $
\rho$}\right] _{ij}  \label{eq:RhoFt}
\end{eqnarray}
where $i,j=1,\ldots ,N$ . This equation can also be derived by taking the
average (defined in eq. (\ref{eq:RhoDef})) of the coherent part (\ref
{eq:SigmaFt}). The contributions which describe the coupling to the leads
are found by adding the average of the incoherent contributions (\ref
{eq:SigmaF-}) and (\ref{eq:SigmaF+}) to the rhs of (\ref{eq:RhoFt}). We
obtain the following closed system of only $N^{2}$ equations which describes
the dynamics of the average occupation number matrix:%
\begin{mathletters}%
%
\begin{eqnarray}
\text{$\partial _{t}\rho $}_{ii} &=&\text{i$\left( t_{i-1}\rho _{ii-1}+\text{%
$t_{i}$}\rho _{ii+1}-\text{$t_{i}$}\rho _{i+1i}-t_{i-1}\rho _{i-1i}\right) $}
\nonumber \\
&&+\text{$\Gamma _{L}\left( 1-\rho _{11}\right) \delta _{i1}$}-\text{$\Gamma
_{R}\rho _{NN}\delta _{iN}$}  \label{eq:RhoFii} \\
\partial _{t}\rho _{ij} &=&\text{i}\left( \varepsilon _{j}-\varepsilon
_{i}\right) \rho _{ij}  \nonumber \\
&&+\text{i}\left( t_{j-1}\rho _{ij-1}+t_{j}\rho _{ij+1}-t_{i-1}\rho
_{i-1j}-t_{i}\rho _{i+1j}\right)  \nonumber \\
&&-\frac{1}{2}\Gamma _{L}\rho _{1j}\delta _{i1}-\frac{1}{2}\Gamma _{R}\rho
_{iN}\delta _{jN}  \label{eq:RhoFij}
\end{eqnarray}
\label{eq:RhoF}%
\end{mathletters}%
%
where $j>i=1,\ldots ,N$ and $t_{0}\equiv t_{N}\equiv 0$. These equations
closely resemble those for the Coulomb blockade case: the coherent
contributions are exactly the same, but in contrast to eq. (\ref{eq:RhoCBii}%
) the average occupations $\rho _{11},\rho _{NN}$ in eq. (\ref{eq:RhoFii})
are not coupled by incoherent transitions to some vacuum state. Furthermore,
in eq. (\ref{eq:RhoFij}) there is a negative contribution due to the
tunneling of an electron {\em into} dot $1$ with rate $\Gamma _{L}$ which is
absent in eq. (\ref{eq:RhoCBij}) due to Coulomb blockade.

Despite the close resemblance to eq. (\ref{eq:RhoCB}) the (calculation of)
stationary solution of eqns. (\ref{eq:RhoF}) for the general case is far
more complicated. An analytical expression seems neither feasible nor
instructive and we consider only a few representative cases here. Assuming
equal couplings $t_{i}=t$ one finds that the resonant current peak ($%
\varepsilon _{ij}=0$) is independent of $N$: 
\begin{equation}
{\displaystyle{I_{N,\text{max}}^{\text{F}} \over e}}%
=\frac{1}{%
{\displaystyle{1 \over \Gamma _{L}}}%
+%
{\displaystyle{1 \over \Gamma _{R}}}%
+%
{\displaystyle{\Gamma _{L} \over 4t^{2}}}%
+%
{\displaystyle{\Gamma _{R} \over 4t^{2}}}%
}  \label{eq:ImaxF}
\end{equation}
This result was previously obtained by Frishman and Gurvitz for tunneling
through multiple-well heterostructures\cite{bib:FrishmanGurvitz} using
essentially the same approach as for the derivation of the modified
Liouville equation\cite{bib:GurvitzPRB57} which we have used (eq. (\ref
{eq:Sigma})). By solving eqns. (\ref{eq:RhoF}) for $N=2$ we reproduce the
result obtained by Gurvitz \cite{bib:GurvitzPRB44},\cite{bib:GurvitzPrager}
from eqns. (\ref{eq:SigmaF}) 
\begin{equation}
\frac{I_{2}^{\text{F}}}{e}=\frac{1}{%
{\displaystyle{1 \over \Gamma _{L}}}%
+%
{\displaystyle{1 \over \Gamma _{R}}}%
+%
{\displaystyle{1 \over \Gamma }}%
\left( 
{\displaystyle{\varepsilon _{12} \over t_{1}}}%
\right) ^{2}+%
{\displaystyle{\Gamma _{L} \over 4t_{1}^{2}}}%
+%
{\displaystyle{\Gamma _{R} \over 4t_{1}^{2}}}%
}  \label{eq:IF2}
\end{equation}
where $\Gamma =\Gamma _{L}+\Gamma _{R}$. For $N=3$ solving the $9$ eqns. (%
\ref{eq:RhoF}) gives the current through a triple dot: 
\begin{equation}
\frac{I_{3}^{\text{F}}}{e}=\frac{1}{\text{$%
{\displaystyle{1 \over \Gamma \text{$_{L}$}}}%
$}+%
{\displaystyle{1 \over \Gamma _{R}}}%
+\text{$%
{\displaystyle{1 \over \Gamma \text{$_{L}$}}}%
$}\left( 
{\displaystyle{\varepsilon _{12} \over t_{1}}}%
\right) ^{2}+%
{\displaystyle{1 \over \Gamma _{R}}}%
\left( 
{\displaystyle{\varepsilon _{23} \over t_{2}}}%
\right) ^{2}+%
{\displaystyle{1 \over \Gamma }}%
\left( 
{\displaystyle{t_{1} \over t_{2}}}%
-%
{\displaystyle{t_{2} \over t_{1}}}%
\right) ^{2}+%
{\displaystyle{\Gamma _{L} \over 4t_{1}^{2}}}%
+%
{\displaystyle{\Gamma _{R} \over 4t_{2}^{2}}}%
-%
{\displaystyle{\left( \frac{1}{\Gamma _{L}}\frac{\varepsilon _{12}}{t_{1}}-\frac{1}{\Gamma _{R}}\frac{t_{1}}{t_{2}}\frac{\varepsilon _{23}}{t_{2}}-\frac{1}{\Gamma }\left( \frac{t_{1}}{t_{2}}-\frac{t_{2}}{t_{1}}\right) \frac{\varepsilon _{13}}{t_{2}}\right) ^{2} \over \frac{1}{\Gamma _{L}}+\frac{1}{\Gamma _{R}}\left( \frac{t_{1}}{t_{2}}\right) ^{2}+\frac{1}{\Gamma }\left( \frac{\varepsilon _{13}}{t_{2}}\right) ^{2}+\frac{\Gamma }{4t_{2}^{2}}}}%
}  \label{eq:IF3}
\end{equation}
If one would first calculate $\mbox{\boldmath $ \sigma $}$ then the subset
of $20$ relevant equations of (\ref{eq:SigmaF}) (containing in total $64$
equations) needs to be solved. For the case where the energies are
configured as a ``Stark ladder'' of total width $\varepsilon $ we have
plotted\ the current in Fig. \ref{fig:IStark}a. Comparison with Fig. \ref
{fig:IStark}b shows that the current for the ``free'' electron case is less
sensitive to localization effects than in the Coulomb blockade case.

\section{Approximation of independent conduction channels}

\label{sec:Appr}It is instructive to consider an approximate approach of
independent conduction channels to our problem. In Section \ref{sec:Weak}
this approximation is introduced assuming that both reservoirs are {\em %
weakly} coupled to dots in the array. In this case the time needed for
tunneling to or from a reservoir $\Gamma ^{-1}$ is much longer than typical
time $t^{-1}$ of the evolution of a coherent state in the array. Therefore
an electron completes many coherent oscillations in the array before
tunneling. Somewhat surprisingly this approach can also be used in the
opposite limit of very {\em strong} coupling to the reservoirs as is shown
in Section \ref{sec:Strong}. In Section \ref{sec:Intermediate} we compare
the results of the independent channel approximation and discuss some
peculiar features of the more general results obtained in \ref{sec:CB} and 
\ref{sec:F}.

\subsection{Weak coupling to the leads}

\label{sec:Weak}We first consider the array of dots without the leads. Let's
denote the $N$ localized (delocalized) eigenstates of a {\em single}
electron in the array of uncoupled (coupled) dots by $|i\rangle $ ($%
|i^{\prime }\rangle $) where $i=1,\ldots ,N$ and the single-electron vacuum
by $|0\rangle $. By transforming to the basis of delocalized states that
diagonalizes $\hat{H}_{0}$ (eq. (\ref{eq:H0})) we obtain new fermionic
operators 
\[
\hat{a}_{i}^{\prime \dagger }=\sum_{k=1}^{N}\langle k|i^{\prime }\rangle 
\hat{a}_{k}^{\dagger },\qquad \hat{a}_{i}^{\prime }=\sum_{k=1}^{N}\langle
^{\prime }i|k\rangle \hat{a}_{k} 
\]
Since in a delocalized state there is a non-zero probability for finding an
electron in both dot $1$ and dot $N$ such a state can be regarded as a {\em %
conduction channel} which carries a current. The new operators add an
electron to channel $|i^{\prime }\rangle $ resp. remove an electron from
channel $|i^{\prime }\rangle $. By expanding the density operator of the
array $\hat{\sigma}$ in the ``many-channel'' basis states $|n_{1}\ldots
n_{N}{}^{\prime }\rangle =|\{n_{k}\}^{\prime }\rangle =\prod_{k=1}^{N}\hat{a}%
_{k}^{\prime \dagger n_{k}}|\{0\rangle $ we obtain a new density matrix $%
\mbox{\boldmath $\sigma$}^{\prime }$. Like in Section \ref{sec:Arrays} we
define an average occupation matrix for the channels $\rho _{ij}^{\prime
}=\langle \hat{a}_{j}^{\prime \dagger }\hat{a}_{i}^{\prime }\rangle $ which
can be expressed in the density matrix $\mbox{\boldmath $\sigma$}^{\prime }$%
. The advantage of the new basis is that the dynamical equations for the
average occupations $\rho _{ii}^{\prime }$ are decoupled from the
non-diagonal elements when we only take $\hat{H}_{0}$ into account: $%
\partial _{t}\rho _{ij}^{\prime }=-$i$(\varepsilon _{i}^{\prime
}-\varepsilon _{j}^{\prime })\rho _{ij}^{\prime }$ where $\varepsilon
_{i,j}^{\prime }$ are the energies of the delocalized states $i,j$. In the
Coulomb blockade case the effect of $\hat{H}_{u}$ is easily translated since
the basis transformation preserves the trace of $\mbox{\boldmath $\rho$}$:
the total occupancy of the channels is restricted to values $\leq $ $1$
whereas in the ``free'' electron case all $N$ channels can be occupied (Fig. 
\ref{fig:NChannels}).

Now we include the coupling to the reservoirs by a Golden Rule approach.
Each channel $|i^{\prime }\rangle $ is connected to lead $L$ with modified
matrix element $t_{L}^{i}=\langle i^{\prime }|1\rangle t_{L}$ and to lead $R$
with $t_{R}^{i}=\langle i^{\prime }|N\rangle t_{R}$: an electron in lead $L$
can tunnel through {\em any} channel to lead $R$. The rate for tunneling
into channel $|i^{\prime }\rangle $ is proportional to the probability to
find the electron in dot $1$whereas the rate for tunneling out of channel $%
|i^{\prime }\rangle $ is proportional to the probability to find the
electron in $N$: 
\begin{equation}
\Gamma _{L}^{i}=\Gamma _{L}\left| \langle i^{\prime }|1\rangle \right|
^{2}\qquad \Gamma _{R}^{i}=\Gamma _{R}\left| \langle i^{\prime }|N\rangle
\right| ^{2}  \label{eq:Gamma[i]}
\end{equation}
These probabilities depend on the energies $\varepsilon _{i}$\ of the
localized states and the matrix elements $t_{i}$ which couple them. When we
assume that the uncertainty in the energy during the tunneling of an
electron between leads and dots is much smaller than the level splitting we
can disregard correlations due to the simultaneous tunneling of\ one
electron through multiple channels. Thus for weak coupling we can include
the leads by only modifying the equations for the {\em diagonal} elements of 
$\mbox{\boldmath $\rho$}^{\prime }$ i.e. the channel occupations. The
channels can be treated as {\em independent} and their contributions to the
current can simply be added.

In the Coulomb blockade case (Fig. \ref{fig:NChannels}a) the interdot
repulsion couples the occupations of the channels and restricts their sum $%
\sum_{i=1}^{N}\rho _{ii}^{\prime }<1$ . The tunneling of an electron into
the empty device with rate $\Gamma _{L}^{i}$ fills a channel $|i^{\prime
}\rangle $ and must tunnel out with rate $\Gamma _{R}^{i}$ before another
electron can tunnels into the device and occupy one channel. The equations
for the occupations are%
\begin{mathletters}%
%
\begin{eqnarray*}
\partial _{t}\rho _{00}^{\prime } &=&\sum_{i=1}^{N}\Gamma _{R}^{i}\rho
_{ii}^{\prime }-(\sum_{i=1}^{N}\Gamma _{L}^{i})\rho _{00}^{\prime } \\
\partial _{t}\rho _{ii}^{\prime } &=&\Gamma _{L}^{i}\rho _{00}^{\prime
}-\Gamma _{R}^{i}\rho _{ii}^{\prime }
\end{eqnarray*}
\end{mathletters}%
%
Using the implied conservation of probability ($\sum_{i=0}^{N}\rho
_{ii}^{\prime }\left( t\right) =1$) the stationary solution ($%
\lim_{t\rightarrow \infty }\partial _{t}\rho _{ii}^{\prime }\left( t\right)
=0$)\ is easily found. Summation of the contributions of the independent
channels gives the stationary current: 
\begin{equation}
\frac{I_{N}^{\text{CB}}}{e}=\sum\limits_{i=1}^{N}\Gamma _{L}^{i}\rho
_{ii}^{\prime }\left( \infty \right) =\frac{\sum\limits_{i=1}^{N}\Gamma
_{L}^{i}}{1+\sum\limits_{i=1}^{N}%
{\displaystyle{\Gamma _{L}^{i} \over \Gamma _{R}^{i}}}%
}=\frac{1}{%
{\displaystyle{1 \over \Gamma _{L}}}%
+%
{\displaystyle{1 \over \Gamma _{R}}}%
F_{N}}  \label{eq:ICB'}
\end{equation}
This is just the expression for the current through one double barrier: the
tunneling time is just the sum of the tunneling times of the individual
barriers where the time for tunneling through the right barrier is increased
by a factor 
\begin{equation}
F_{N}=\sum_{i=1}^{N}\frac{\left| \langle i^{\prime }|1\rangle \right| ^{2}}{%
\left| \langle i^{\prime }|N\rangle \right| ^{2}}>\frac{\sum%
\limits_{i=1}^{N}\left| \langle i^{\prime }|1\rangle \right| ^{2}}{%
\sum\limits_{i=1}^{N}\left| \langle i^{\prime }|N\rangle \right| ^{2}}=1
\label{eq:F[N]Weak}
\end{equation}

In the ``free'' electron case (Fig. \ref{fig:NChannels}b) the occupations of
the channels are not coupled and their sum is restricted only by $%
\sum_{i=1}^{N}\rho _{ii}^{\prime }<N$ . The device is equivalent to $N$
independent double barriers in parallel where the occupancy of channel $i$
obeys 
\[
\partial _{t}\rho _{ii}^{\prime }=\Gamma _{L}^{i}\left( 1-\rho _{ii}^{\prime
}\right) -\Gamma _{R}^{i}\rho _{ii}^{\prime } 
\]
From the stationary solution ($\lim_{t\rightarrow \infty }\partial _{t}\rho
_{ii}^{\prime }\left( t\right) =0$)\ the current is readily found: 
\begin{equation}
\frac{I_{N}^{\text{F}}}{e}=\sum_{i=1}^{N}\Gamma _{L}^{i}\rho _{ii}^{\prime
}\left( \infty \right) =\sum_{i=1}^{N}\frac{1}{%
{\displaystyle{1 \over \Gamma _{L}^{i}}}%
+%
{\displaystyle{1 \over \Gamma _{R}^{i}}}%
}  \label{eq:IF'}
\end{equation}
The tunneling time through each double barrier is the sum of the tunneling
times $\left( \Gamma _{L,R}^{i}\right) ^{-1}>\Gamma _{L,R}^{-1}$ whereas the
tunnel rate of the device is the sum of the tunnel rates of the $N$ double
barriers.

For a few representative cases we have explicitly worked out the approximate
approach in the limit of weak coupling to both leads $\Gamma _{L,R}\ll
\varepsilon _{ij},t_{i}$. For the case of $N=2$ dots we have first
calculated the $2$ exact eigenstates of one electron in the array of coupled
dots without the reservoirs. From these we obtained the tunnel rates to and
from each of the delocalized states. Finally we summed the contributions of
the independent channels to the total current: eq. (\ref{eq:ICB'}) precisely
gives the result (\ref{eq:ICB2}) without the term $\left( \Gamma
_{R}/2t_{1}\right) ^{2}$ whereas (\ref{eq:IF'}) gives the result (\ref
{eq:IF2}) without the term $\Gamma _{L}\Gamma _{R}/\left( 2t_{1}\right) ^{2}$
(The latter result was previously found in \cite{bib:KorotkovAverinLikharev}
for the case of a double quantum well). For the case of $N$\ dots with equal
interdot couplings $t_{i}=t$\ and aligned levels $\varepsilon
_{i}=\varepsilon _{j}$ a simple result is also possible because in each
delocalized state $|i^{\prime }\rangle $ the electronic densities in the
dots are spatially symmetric. Then the modification factors of the rates are
equal for each channel ($\left| \langle i^{\prime }|1\rangle \right|
^{2}=\left| \langle i^{\prime }|N\rangle \right| ^{2}$) and they cancel in (%
\ref{eq:ICB'}) and (\ref{eq:IF'}). Thus for weak coupling to both reservoirs
we obtain 
\begin{eqnarray}
\frac{I_{N,\text{max}}^{\text{CB}}}{e} &=&\frac{1}{%
{\displaystyle{1 \over \Gamma _{L}}}%
+%
{\displaystyle{1 \over \Gamma _{R}}}%
N}  \label{eq:ImaxCBweak} \\
\frac{I_{N,\text{max}}^{\text{F}}}{e} &=&\frac{1}{%
{\displaystyle{1 \over \Gamma _{L}}}%
+%
{\displaystyle{1 \over \Gamma _{R}}}%
}  \label{eq:ImaxFweak}
\end{eqnarray}
which is just (\ref{eq:ImaxCB}) and (\ref{eq:ImaxF}) without the terms $%
\left( N-1\right) \left( \Gamma _{R}/2t\right) ^{2}$ and $\Gamma _{L}\Gamma
_{R}/\left( 2t\right) ^{2}$ respectively$.$ Coulomb blockade increases the 
{\em effective} time for tunneling out of the device by a factor $N$ w.r.t.
the ``free'' electron case. The current increases {\em linearly} with each
rate $\Gamma _{L},\Gamma _{R}\ll t$ as expected from the enhanced tunneling.

\subsection{Strong coupling to the leads}

\label{sec:Strong}The terms which were found missing above become important
when one of the tunnel rates to the reservoirs is comparable to the coherent
interdot coupling. The correlations between the conduction channels can no
longer be disregarded in this case: eqns. (\ref{eq:RhoCB}) resp. (\ref
{eq:RhoF}) do not decouple into separate sets of equations for diagonal and
non-diagonal elements on the basis of delocalized states. However, we will
now show that in the limit of {\em strong} coupling to one or both of the
reservoirs we can use the independent channel approximation again. (Note
that the energy uncertainties are assumed to be smaller than the bias $%
\Gamma _{L,R}\ll \mu _{L}-\mu _{R}$)

First we consider the case where only the last dot $N$ is strongly coupled
to the right lead: $\Gamma _{L}\ll \varepsilon _{ij},t_{i}\ll \Gamma _{R}$.
The electronic state of the dot $N$ and reservoir $R$ form a continuum of
states with a Lorentzian spectral density which is approximately constant
over the energy range $t_{N-1}\ll \Gamma _{R}$: 
\[
D_{N+R}\left( \varepsilon \right) =\frac{1}{2\pi }\frac{\Gamma _{R}}{\left(
\varepsilon -\varepsilon _{N}\right) ^{2}+\left( \Gamma _{R}/2\right) ^{2}}%
\approx \frac{1}{2\pi }\frac{4}{\Gamma _{R}} 
\]
The array of the remaining $N-1$ dots is {\em weakly} coupled to this
continuum of states with matrix element $t_{N-1}$. Therefore we can apply
the independent channel approach: the tunnel rate from dot $N-1$ to the
continuum on the right is found with the Golden Rule 
\begin{equation}
\tilde{\Gamma}_{R}=2\pi t_{N-1}^{2}D_{N+R}\left( \varepsilon \right) =\frac{%
4t_{N-1}^{2}}{\Gamma _{R}}  \label{eq:Gamma[R]Tilde}
\end{equation}
For the case of aligned levels $\varepsilon _{ij}=0$ and equal couplings $%
t_{i}=t$ substitution of $\Gamma _{R}\rightarrow \tilde{\Gamma}_{R}$ and $%
N\rightarrow N-1$ in (\ref{eq:ImaxCBweak}) resp. (\ref{eq:ImaxFweak}) gives
the maximum current in the limit of strong coupling to the right lead: 
\begin{eqnarray}
\frac{I_{N,\text{max}}^{\text{CB}}}{e} &=&\frac{1}{%
{\displaystyle{1 \over \Gamma _{L}}}%
+%
{\displaystyle{N-1 \over \tilde{\Gamma}_{R}}}%
}=\frac{1}{%
{\displaystyle{1 \over \Gamma _{L}}}%
+%
{\displaystyle{\Gamma _{R} \over 4t^{2}}}%
\left( N-1\right) }  \label{eq:ImaxCBstrong} \\
\frac{I_{N,\text{max}}^{\text{F}}}{e} &=&\frac{1}{%
{\displaystyle{1 \over \Gamma _{L}}}%
+%
{\displaystyle{1 \over \tilde{\Gamma}_{R}}}%
}=\frac{1}{%
{\displaystyle{1 \over \Gamma _{L}}}%
+%
{\displaystyle{\Gamma _{R} \over 4t^{2}}}%
}  \label{eq:ImaxFstrong}
\end{eqnarray}
which is in agreement with our result (\ref{eq:ImaxCB})\ resp. (\ref
{eq:ImaxF}). For strong coupling to the right lead the current {\em decreases%
} as $\Gamma _{R}^{-1}$ which is somewhat surprising because tunneling is
enhanced. The origin of this effect is the formation of linear combinations
of the discrete state in dot $N$ with reservoir states with energies roughly
between $\varepsilon _{N}\pm \Gamma _{R}$. Because the tunnel processes from
reservoir states back into the discrete state destructively interfere\cite
{bib:Merzbacher}, the discrete state irreversibly decays into the continuum.
The resulting spectral density decreases with the energy uncertainty $\Gamma
_{R}$. The eventual decrease of the current with the tunnel rate $\Gamma
_{R} $ can also be interpreted as a manifestation of the quantum Zeno effect
as discussed in \cite{bib:GurvitzPRB57}. More general we have the following
relation: for $\Gamma _{L}\ll \varepsilon _{ij},t_{i}\ll \Gamma _{R}$ 
\begin{eqnarray*}
I_{N}^{\text{CB}}\left( \varepsilon _{1}\ldots \varepsilon _{N},t_{1}\ldots
t_{N-1};\Gamma _{L},\Gamma _{R}\right) &=&I_{N-1}^{\text{CB}}(\varepsilon
_{1}\ldots \varepsilon _{N-1},t_{1}\ldots t_{N-2};\Gamma _{L},\frac{%
4t_{N-1}^{2}}{\Gamma _{R}}) \\
I_{N}^{\text{F}}\left( \varepsilon _{1}\ldots \varepsilon _{N},t_{1}\ldots
t_{N-1};\Gamma _{L},\Gamma _{R}\right) &=&I_{N-1}^{\text{F}}(\varepsilon
_{1}\ldots \varepsilon _{N-1},t_{1}\ldots t_{N-2};\Gamma _{L},\frac{%
4t_{N-1}^{2}}{\Gamma _{R}})
\end{eqnarray*}
This is clearly satisfied by (\ref{eq:ICB3}) resp. and (\ref{eq:IF3}) and
the general form (\ref{eq:ICB}) has this property.

Now consider the case where the array is coupled strongly to both leads i.e. 
$\varepsilon _{ij},t_{i}\ll \Gamma _{L,R}$. In the ``free'' electron case
the reservoir $L$ coupled to dot $1$ gives a new continuum of states with a
spectral density which is approximately constant over an energy range $%
t_{1}\ll \Gamma _{L}$: 
\[
D_{1+L}\left( \varepsilon \right) =\frac{1}{2\pi }\frac{\Gamma _{L}}{\left(
\varepsilon -\varepsilon _{1}\right) ^{2}+\left( \Gamma _{L}/2\right) ^{2}}%
\approx \frac{1}{2\pi }\frac{4}{\Gamma _{L}} 
\]
The tunnel rate from this continuum to dot $2$ is 
\[
\tilde{\Gamma}_{L}=2\pi t_{1}^{2}D_{1+L}\left( \varepsilon \right) =\frac{%
4t_{1}^{2}}{\Gamma _{L}} 
\]
The remaining $N-2$ dots are weakly coupled to a continuum on the left with
matrix element $t_{1}$ and to the right with $t_{N-1}$ and we can applying
the independent channel approach to the $N-2$ conduction channels. For the
case of aligned levels $\varepsilon _{i}=\varepsilon _{j}$ and equal
couplings $t_{i}=t$ substitution of both $\Gamma _{L}\rightarrow \tilde{%
\Gamma}_{L}$ and $\Gamma _{R}\rightarrow \tilde{\Gamma}_{R}$ in (\ref
{eq:ImaxFweak}) gives the maximum current in the limit of strong coupling to
both leads 
\[
\frac{I_{\text{max}}^{\text{F}}}{e}=\frac{1}{%
{\displaystyle{1 \over \tilde{\Gamma}_{L}}}%
+%
{\displaystyle{1 \over \tilde{\Gamma}_{R}}}%
}=\frac{1}{%
{\displaystyle{\Gamma _{L}+\Gamma _{R} \over 4t^{2}}}%
} 
\]
in agreement with result (\ref{eq:ImaxF}). More general we have the relation
for $\varepsilon _{i},t_{i}\ll \Gamma _{L,R}$: 
\[
I_{N}^{\text{F}}\left( \varepsilon _{1}\ldots \varepsilon _{N},t_{1}\ldots
t_{N-1};\Gamma _{L},\Gamma _{R}\right) =I_{N-2}^{\text{F}}(\varepsilon
_{2}\ldots \varepsilon _{N-2},t_{2}\ldots t_{N-2};\frac{4t_{1}^{2}}{\Gamma
_{L}},\frac{4t_{N-1}^{2}}{\Gamma _{R}}) 
\]
which is satisfied by (\ref{eq:IF3}). In the Coulomb blockade case interdot
repulsion prevents the discrete state in dot $1$ from mixing with the
reservoir: even if the energy uncertainty allows tunneling into dot $1$ to
occur on a small time scale $\Gamma _{L}^{-1}$, the next electron will have
to wait for the previous one to tunnel out which occurs on the much larger
time scale $\max \{\varepsilon _{ij}^{-1},t^{-1},\Gamma _{R}^{-1}\}$. In the
limit of strong coupling to both leads the current in this case is {\em %
independent} of $\Gamma _{L}$ and is correctly given by (\ref
{eq:ImaxCBstrong}) in agreement with result (\ref{eq:ImaxCB}).

Finally, we consider the case where only dot $1$ is strongly coupled to the
left lead i.e. $\Gamma _{R}\ll \varepsilon _{ij},t_{i}\ll \Gamma _{L}$. In
the Coulomb blockade case the weak coupling result still applies as
explained above. In the ``free'' electron case the discussion is completely
analogous to the case of strong coupling of dot $N$ to the right lead and
the result is obtained by simply interchanging $L\leftrightarrow R$ and $%
1\leftrightarrow N$..

\subsection{Intermediate coupling to the leads; maximum current}

\label{sec:Intermediate}The competition between enhanced tunneling for weak
coupling to the leads and destructive interference in the opposite limit
implies that the current reaches a {\em maximum} value when the rate for
tunneling into a dot is comparable to the coherent coupling to the
neighboring dot. We can find the precise location and value of this maximum
with the results obtained in Section \ref{sec:CB} and \ref{sec:F} which also
hold in this intermediate case.

First we consider the resonant current peak (\ref{eq:ImaxCB}) and (\ref
{eq:ImaxF}) as a function of the transparency $\Gamma _{R}$ of the right
tunnel barrier as plotted in Fig. \ref{fig:Imax}. Starting from zero the
current initially increases linearly as expected from the enhanced tunneling
to the right lead. Then a maximum is reached:%
\begin{mathletters}%
%

\begin{eqnarray}
\frac{I_{\text{max}}^{\text{CB}}}{e} &=&%
{\displaystyle{1 \over %
{\displaystyle{1 \over \Gamma _{L}}}+\sqrt{N\left( N-1\right) }%
{\displaystyle{1 \over t}}}}%
,\quad \Gamma _{R}=\sqrt{\frac{N}{N-1}}2t  \label{eq:ImaxCBmax} \\
\frac{I_{\text{max}}^{\text{F}}}{e} &=&%
{\displaystyle{1 \over %
{\displaystyle{1 \over \Gamma _{L}}}+%
{\displaystyle{\Gamma _{L} \over 4t^{2}}}+%
{\displaystyle{1 \over t}}}}%
,\quad \quad \quad \quad \Gamma _{R}=2t  \label{eq:ImaxFmax}
\end{eqnarray}
\end{mathletters}%
%
Increasing the transparency further will {\em reduce} the current as
explained above. The maximum occurs when there is an optimal balance of the
coherent tunneling into dot $N$ and incoherent tunneling from this dot to
reservoir $R.$ At this point the {\em effective} time for tunneling out of
the device is the same for weak (eq. (\ref{eq:ImaxCBweak}) resp. (\ref
{eq:ImaxFweak})) and strong coupling to the right lead (eq. (\ref
{eq:ImaxCBstrong}) resp. (\ref{eq:ImaxFstrong})) i.e. at these values of $%
\Gamma _{R}$ we have 
\begin{eqnarray*}
\frac{1}{\Gamma _{R}}N &=&\frac{1}{\tilde{\Gamma}_{R}}\left( N-1\right)
\quad \text{ (CB)} \\
\frac{1}{\Gamma _{R}} &=&\frac{1}{\tilde{\Gamma}_{R}}\quad \quad \quad \quad
\quad \text{\ (F)}
\end{eqnarray*}
where $\tilde{\Gamma}_{R}$ is given by (\ref{eq:Gamma[R]Tilde}). For the
case of ``free'' electrons the effective time for tunneling out and
therefore the position of the maximum is independent of the number of dots $N
$. (For this case the $\Gamma _{R}$ dependence of the resonant current peak
has been discussed for a double dot system in \cite{bib:GurvitzPRB57}). In
the Coulomb blockade case the non-monotonic variation of the current with $%
\Gamma _{R}$ persists for all $N$. The maximum occurs at a slightly higher
value of $\Gamma _{R}$ than for the ``free'' electron case: at $\Gamma
_{R}=2t=\tilde{\Gamma}_{R}$ the effective tunneling time for weak coupling
is still larger than for strong coupling because the fraction of channels
excluded by Coulomb repulsion is larger for $N$ dots then for $N-1$ dots.
The tunnel rate must be increased by a factor $\sqrt{N/(N-1)}$ to exactly
balance the effective tunneling times. For large $N$ this difference becomes
negligible and the maximum occurs at the same position as for the ``free''
electron case but with a much smaller amplitude (Fig. \ref{fig:Imax}).

Next we consider the resonant current peak (\ref{eq:ImaxCB}) and (\ref
{eq:ImaxF}) as a function of the transparency $\Gamma _{L}$ of the left
tunnel barrier. In the ``free'' electron case the maximum current (\ref
{eq:ImaxF}) remains unchanged when we interchange $\Gamma _{L}$ and $\Gamma
_{R}$. Therefore the resonant current peak (\ref{eq:ImaxF}) also displays a
maximum as a function of $\Gamma _{L}$ at $\Gamma _{L}=2t$. As a function of
both rates the maximum current is 
\[
\frac{I_{N,\text{max}}^{\text{F}}}{e}=\frac{t}{2},\qquad \Gamma _{L}=\Gamma
_{R}=2t 
\]
As discussed in Section \ref{sec:Strong} there is no such effect in the
Coulomb blockade case: due to interdot repulsion the resonant current peak (%
\ref{eq:ImaxCBmax}) will increase with $\Gamma _{L}$ and saturate at a
maximal value when $\Gamma _{L}\gg t$%
\[
\frac{I_{N,\text{max}}^{\text{CB}}}{e}=\frac{t}{\sqrt{N\left( N-1\right) }}%
,\qquad \Gamma _{L}\gg \Gamma _{R}=\sqrt{\frac{N}{N-1}}2t 
\]
For $N=2$ the maximal current as a function of the rates is larger in the
Coulomb blockade case whereas for $N>2$ it is larger in the ``free''
electron case.

\section{Conclusions}

\label{sec:Concl}We have extended the density matrix approach to resonant
tunneling to the case of a linear array of quantum dots with strong, long
range electron-electron interaction. We have found exact analytical
expressions for the stationary current in the array{\bf \ }for an arbitrary
set of parameters (within the applicability of our model) characterizing the
array. Coulomb repulsion was found to reduce the resonant current by a
factor of the order of the number of dots. The formation of a localized
state in one of the dots when the energy level is displaced results in an
exponential decay of the current with increasing size of the array. Our
approach takes into account correlations between conduction channels in the
array due to the coupling to the electron reservoirs. This makes our results
also valid for relatively strong tunnel coupling to the reservoirs where the
independent channel approximation does not work. These correlations
manifests themselves in the eventual decrease of the resonant current when
the rate for tunneling into the reservoir is increased.

\section{Acknowledgements}

The authors acknowledge valuable discussions with T. H. Stoof, G. E. W.
Bauer and especially S. A. Gurvitz. This work was supported by the Dutch
Foundation for Fundamental Research on Matter (FOM) and by the NEDO joint
research program (NTDP -98).

\begin{figure}[tbp]
\caption{Linear array of $N$ quantum dots coupled to leads $L$ and $R$. The
energy levels of the uncoupled dots are given by full lines when relevant
for resonant transport and dashed lines when irrelevant.}
\label{fig:Ndots}
\end{figure}

\begin{figure}[tbp]
\caption{Normalized resonant current for the Coulomb blockade case with $%
N=2,3,4,5,6$ (full curves downwards) and $N=\infty $ (thick full curve). (a)
Variation of the last last level i.e. $\protect\varepsilon _{i}=\protect%
\varepsilon _{N}\protect\delta _{iN}$ . Increasing $N$ does not alter the
Lorentzian form of the curve. (b) Variation of the first level i.e. $\protect%
\varepsilon _{i}=\protect\varepsilon _{1}\protect\delta _{i1}$. Electrons
are localized in the first dot after tunneling through the left barrier
resulting in the exponential decay of the current tails with $N$. For $N=2$
the curves in (a) and (b) coincide.}
\label{fig:ICB}
\end{figure}

\begin{figure}[tbp]
\caption{Normalized current through an array of $N=2,3,4,5,6$ dots with
energies configured as a Stark ladder $\protect\varepsilon _{i}=i/\left(
N-1\right) \protect\varepsilon $ of varying width $\protect\varepsilon $.
(a) ``Free'' electron case: the curves for $N=2,3$ (indicated by the arrow)
coincide for an isotropic array $t_{i}=t$ with $\Gamma _{L}=\Gamma _{R}$.
(b) Coulomb blockade case. }
\label{fig:IStark}
\end{figure}

\begin{figure}[tbp]
\caption{Maximum resonant current as a function of the transparency of the
right barrier: Coulomb blockade case for $N=1,2,3,4,5,6$ (solid lines
downwards) and ``free'' electron case (dot-dashed line for any $N>1$). The
dotted lines show the position of the maxima.}
\label{fig:Imax}
\end{figure}

\begin{figure}[tbp]
\caption{Tunneling through $N$ independent channels. (a) Coulomb blockade
case: an electron tunneling through one of the channels blocks the remaining 
$N-1$ channels. The state must first decay to the vacuum before another
electron can enter one of the channels. (b) ``Free'' electron case: $N$
parallel channels for tunneling are available.}
\label{fig:NChannels}
\end{figure}


\begin{references}
\bibitem{bib:SpinDegeneracy}  When a very large infinite on-site Coulomb
repulsion is assumed the electron spin can be included by taking into
account the spin-degeneracy of the single dot levels. In a high magnetic
field where all electron spins are aligned this degeneracy is absent and our
results apply directly.
\end{references}
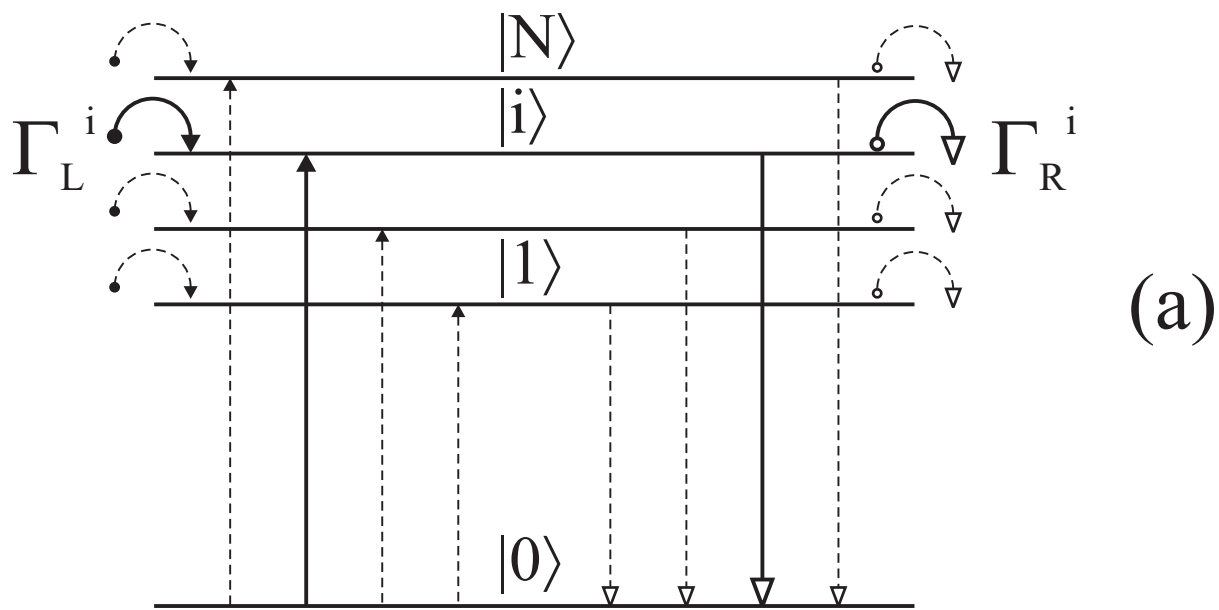
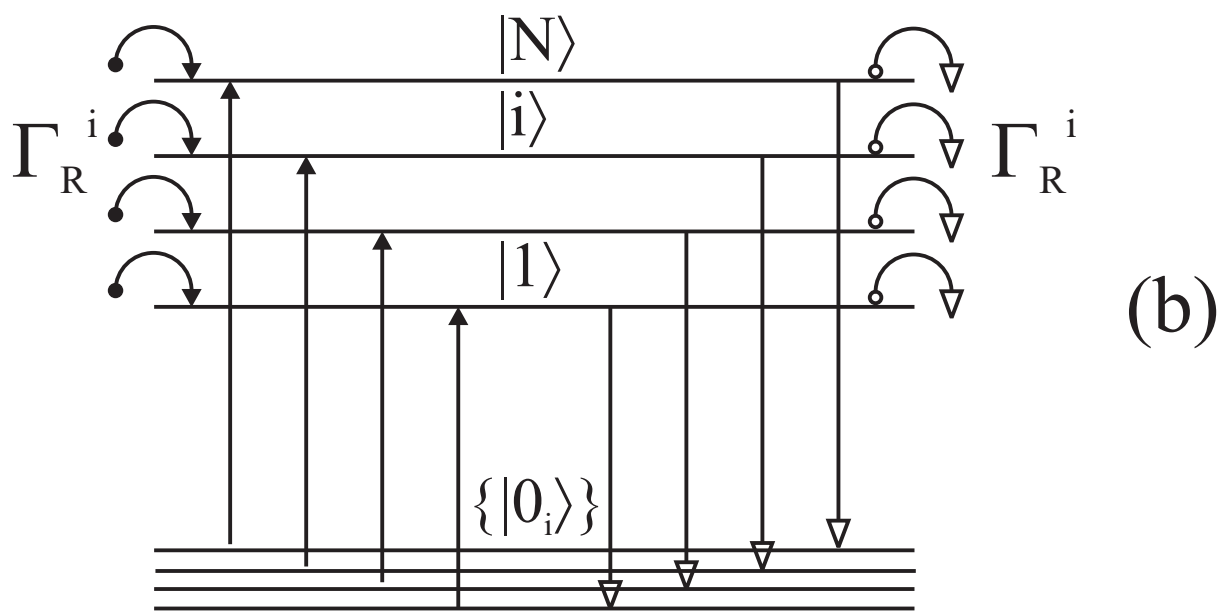
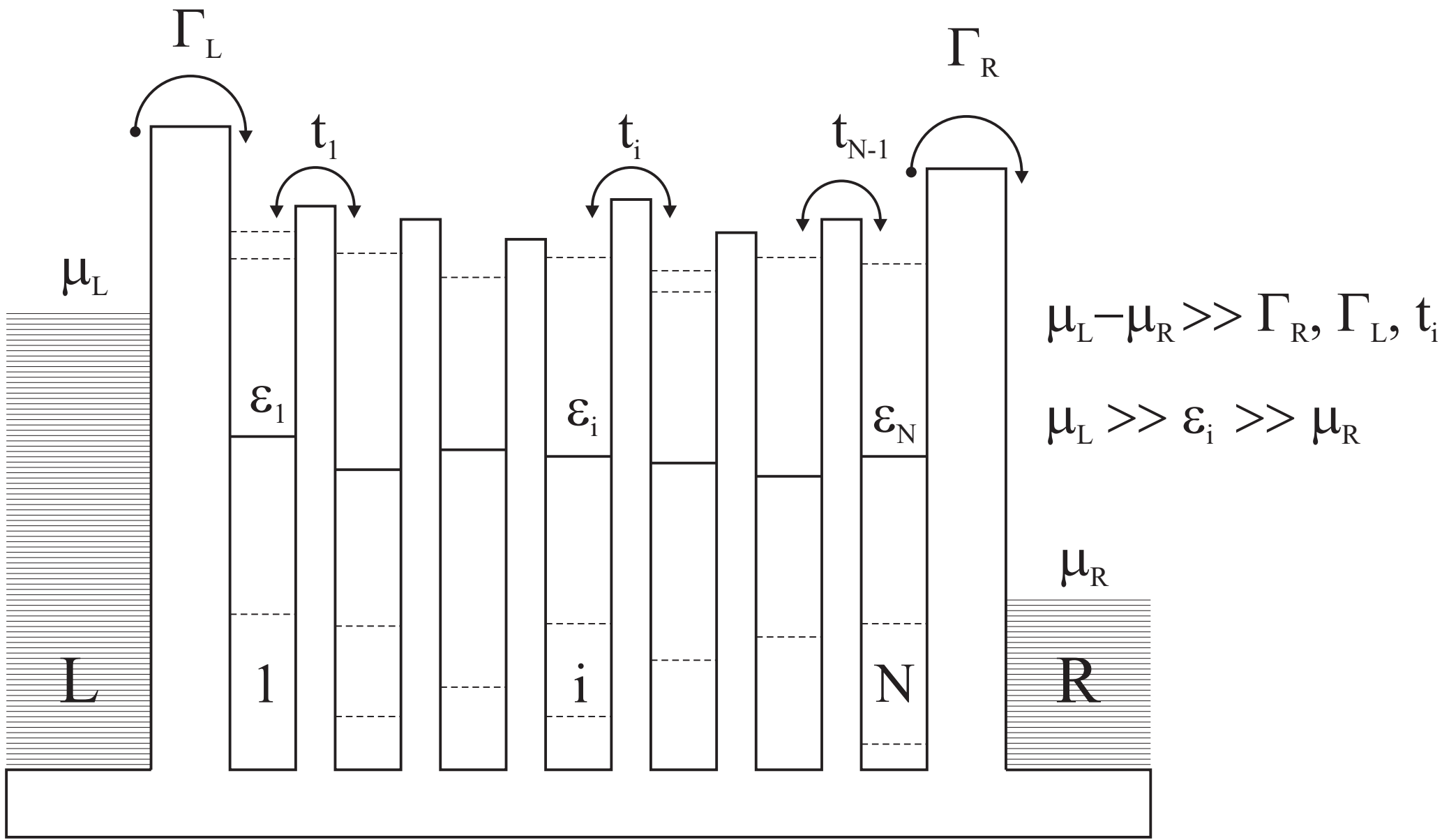
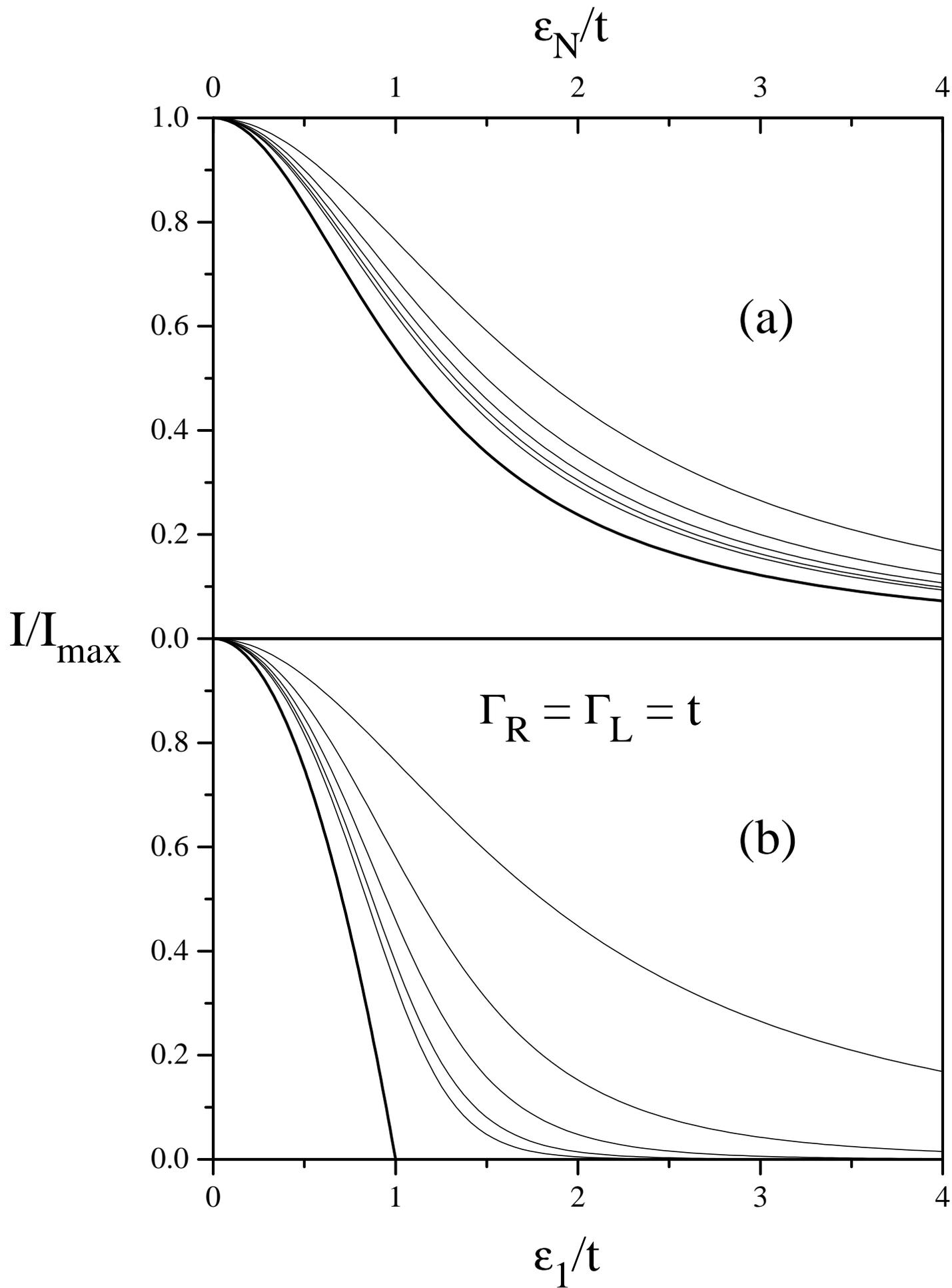
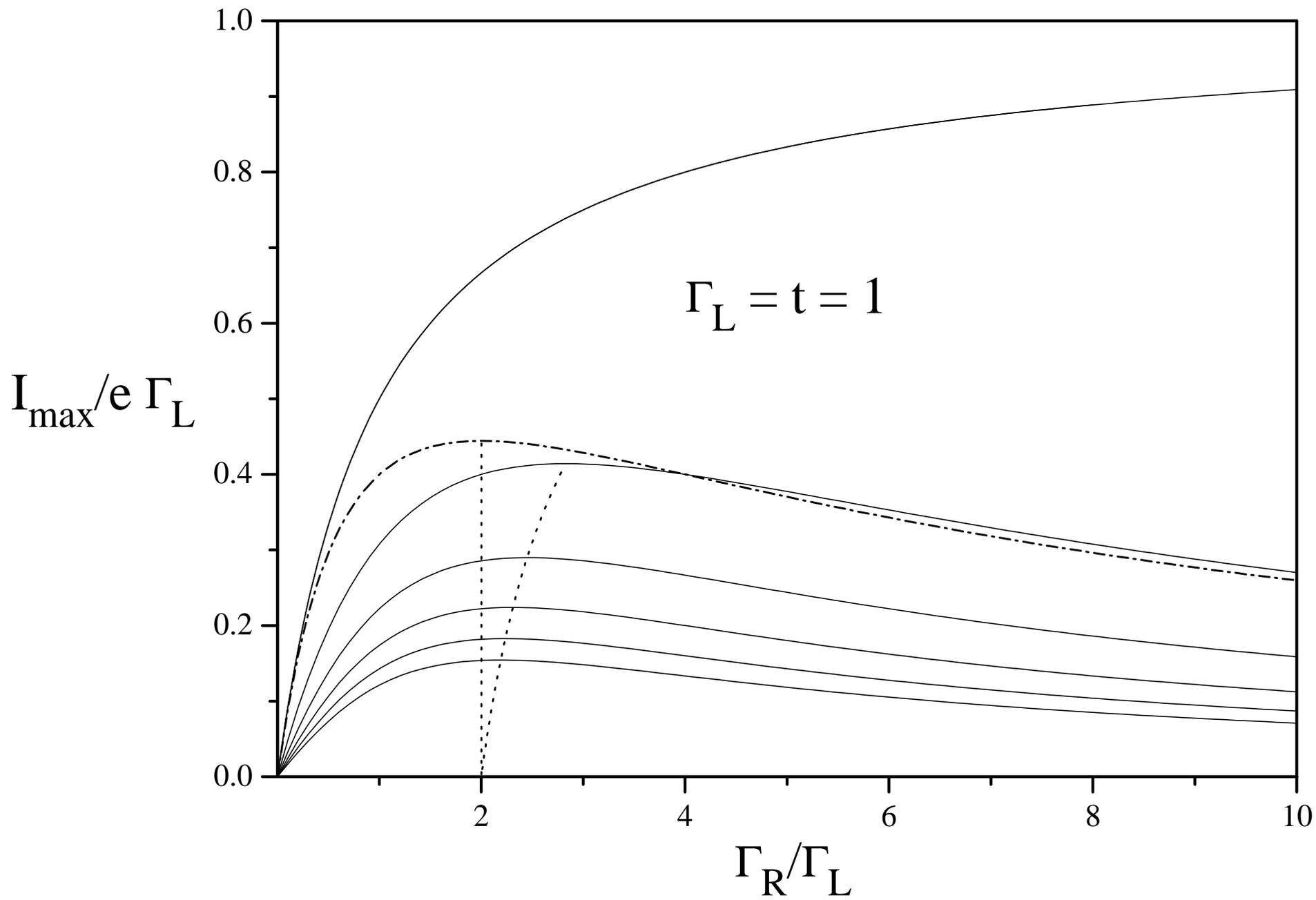
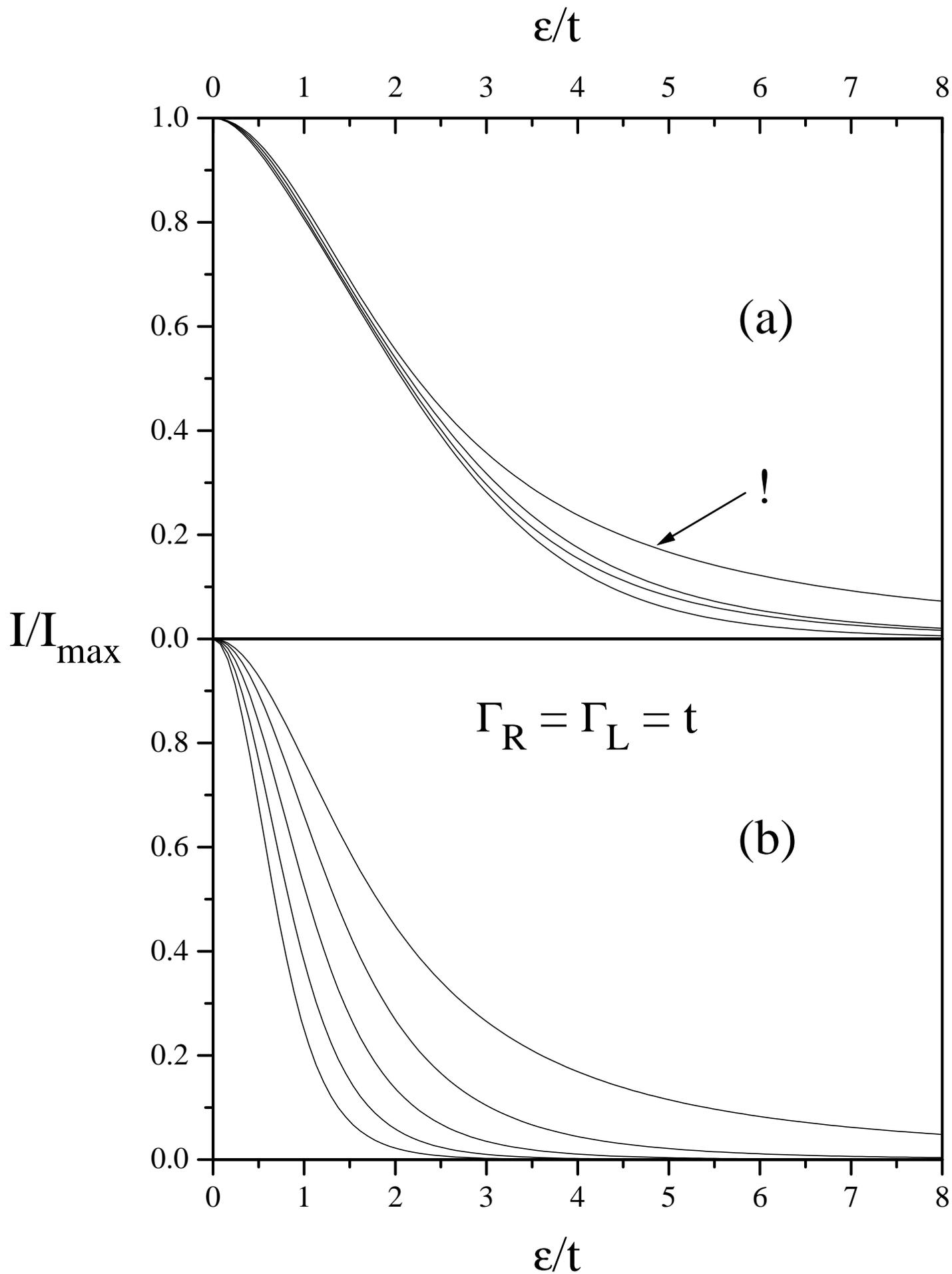
\end{document}